\documentclass[twocolumn,showpacs,amsmath,amssymb]{revtex4}
\usepackage{graphics}
\usepackage{dcolumn}
\usepackage{bm}

\begin{document}


\title{Structure and peculiarities of the (8$\,\bf\times\,${\em{n}})-type  Si(001) surface\\ prepared in a molecular beam epitaxy chamber:\\ a scanning tunneling microscopy study}

\author{Larisa V. Arapkina}
\email{arapkina@kapella.gpi.ru}
\author{Vladimir M. Shevlyuga}%
 \author{Vladimir A. Yuryev}%
\homepage{http://www.gpi.ru/eng/staff_s.php?eng=1&id=125}%
\affiliation{A.\,M.\,Prokhorov General Physics Institute of the Russian Academy of Sciences,\\ 38 Vavilov Street, Moscow, 119991, Russia}%

\date{ }

\begin{abstract}
A clean Si(001) surface thermally purified in the ultrahigh vacuum molecular beam epitaxy chamber has been investigated by means of the scanning tunneling microscopy. The morphological peculiarities of the Si(001) surface have been explored in detail. The classification of surface structure elements has been carried out, the dimensions of the elements have been measured, and relative heights of the surface relief have been determined.  A reconstruction of the Si(001) surface prepared in the molecular-beam epitaxy chamber has been found to be $(8\times n)$. A model of the Si$(001)-(8\times n)$ surface structure is proposed.
\end{abstract}

\maketitle


Investigations of clean silicon surfaces prepared in conditions of actual technological chambers are of great interest due to the requirements of the coming industry which starts to operate on nanometer and subnanometer scale when designing advanced solutions for future nanoelectronic devices.\cite{Report_01-303} The  ``design rules''  of such devices in the nearest future will become comparable with the dimensions of structure features of Si(001) surface, at  least of its high-order reconstructions such as $c(8\times 8)$. Most of  researches of the Si(001) surface have so far been carried out in specially refined conditions which allowed one to study the most common types of the surface reconstructions.\cite{Hamers,Chadi,Hu,Murray,Kubo,Iton,Koo,Hata,Liu1,Okada,Goryachko,Liu2}  The ambient in technological vessels such as molecular beam epitaxy (MBE) chambers  is usually not so pure as in specially refined  ones designed for surface studies---there are many sources of surface contaminants in the process chambers, and construction materials of heaters and evaporators as well as foreign matters used for epitaxy and doping are among them.

In the present paper the Si(001) surface was treated following a standard procedure of Si wafer preparation for the MBE growth.


The experiments were made using an integrated UHV system based on the Riber EVA~32 molecular beam epitaxy chamber  coupled through a transfer line with the GPI~300 scanning tunneling microscope
(STM). This instrument enables the STM study of samples at any stage of Si surface cleaning and MBE growth. The samples can be moved in the STM chamber and back in the MBE chamber never leaving the UHV ambient.

\begin{figure}[b]
\includegraphics{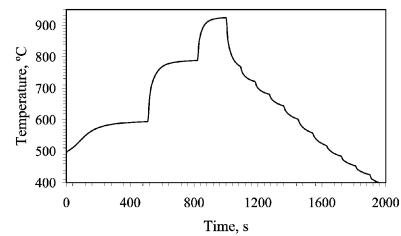}
\caption{\label{fig:cycle} A diagram of the final thermal treatment of samples.}
\end{figure}

The samples were 8$\times$8 mm$^{2}$ squares cut from B-doped    CZ Si$(100)$ wafers (\,$p$-type,  $\rho\,= 12~\Omega\,$cm). After washing and chemical treatment following a standard procedure\cite{Report_01-303} 
the samples  were mounted on the Mo STM holder and clamped with the Ta fasteners. Then they were loaded into the  airlock and transferred to the preliminary annealing chamber where outgassed at  $\sim 565\,^\circ$C and  $\sim 5\times 10^{-9}$ Torr for about 24 hours. After that the samples were moved  for final treatment (Fig.~\ref{fig:cycle}) into the MBE chamber evacuated down to $\sim 10^{-11}$ Torr. There were two shelves at $\sim 600\,^\circ$C ($\sim 5$ min.) and $\sim 800\,^\circ$C ($\sim 3$ min.) at the heating stage.  The annealing at $\sim 900\,^\circ$C took $\sim 2.5$ min. Then the temperature rapidly lowered to $\sim 750\,^\circ$C. The rate of the following cooling down was  $\sim 0.4\,^\circ$C/s. The pressure in the MBE chamber  grew up to $\sim 2\times 10^{-9}$ Torr during the process. 

\begin{figure*}
\includegraphics{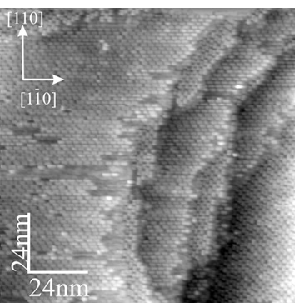} (a) 
\includegraphics{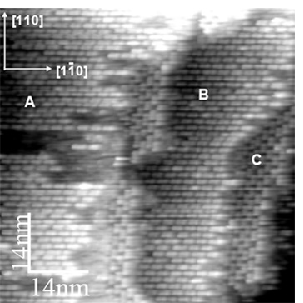} (b) 
\includegraphics{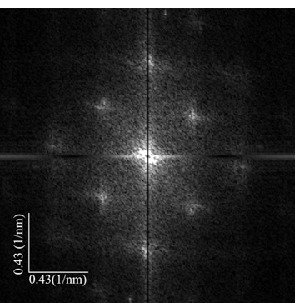} (c)
\caption{\label{fig:view}STM images of the Si(001) surface after annealing cycle shown in Fig.~\ref{fig:cycle} (+1.9~V, 70~pA (a), and +1.9~V, 50~pA (b)), rows run along $[110]$ or $[1\overline{1}0]$ directions; 2-D Fourier transform pattern (c) of the picture (a).}
\end{figure*}

\begin{figure*}
\includegraphics{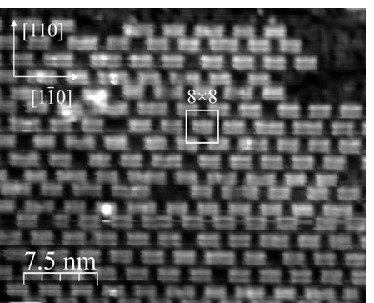} (a)
\includegraphics{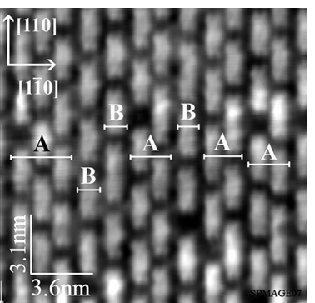} (b)
\includegraphics{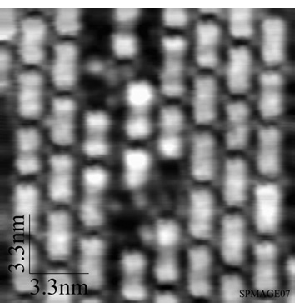} (c)
\caption{\label{fig:8x8}Empty state images of the Si(001) surface (+2.0~V, 200~pA (a), +2.0~V, 150~pA (b),  +1.6~V, 100~pA (c)); a $c(8\times 8)$ unit cell is marked by a white box in (a);    $(8\times 6)$ and  $(8\times 8)$ structures are marked as A and B in (b); row wedging between two rows and the lost blocks are seen in (c).}
\end{figure*}

The samples were heated by  Ta  radiators from the rear side  in both chambers. The temperature was monitored with chromel-allimel and tungsten-rhenium thermocouples in the preliminary annealing and MBE chambers, respectively. The thermocouples were  mounted near the rear side of the samples and {\it in situ} graduated against the IMPAC~IS\,12-Si pyrometer, which measured the Si sample temperature through chamber windows.

\begin{figure*}
\includegraphics{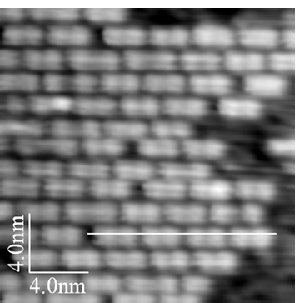} (a)
\includegraphics{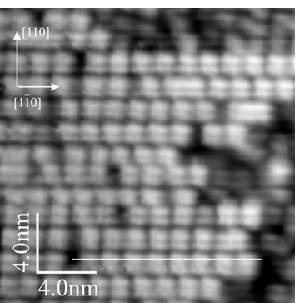} (b)
\includegraphics{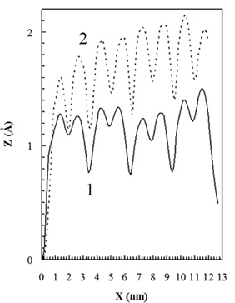} (c)
\caption{\label{fig:empty-filled}Empty (a) and filled (b) state  views of the same region on Si(001) surface (+1.7~V, 100~pA and $-2.0$~V, 100~pA). Positions of line scan profile (c) extremes for empty (1) and filled (2) state distributions along corresponding lines match exactly.}
\end{figure*}

\begin{figure}
\includegraphics{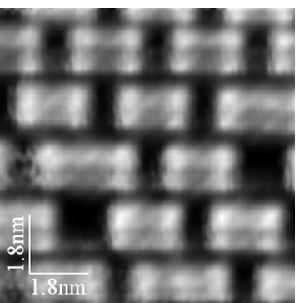} (a)
\includegraphics{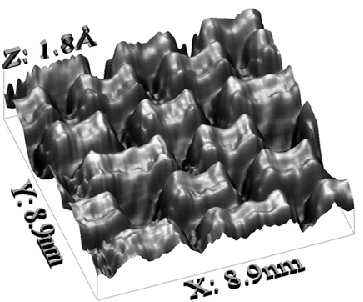} (b)
\caption{\label{fig:close-up}An example of local disordering: 2-D  (a)  and 3-D (b) views (+2.0~V, 200~pA).  Two types of the blocks and period violations forming vacancies are shown.}
\end{figure}

\begin{figure}
\includegraphics{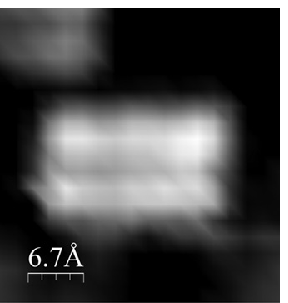} (a)
\includegraphics{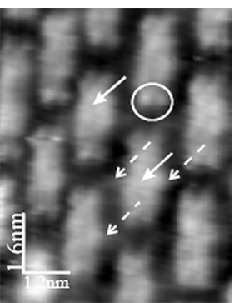} (b)
\caption{\label{fig:block}An empty state view of the short block (a) (+2.0~V, 200~pA) and the rows (b) (+2.0~V, 150~pA).}
\end{figure}

\begin{figure}
\includegraphics{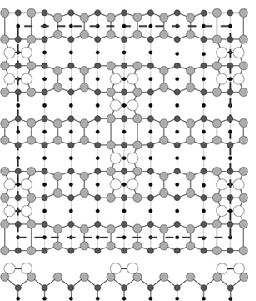} (a)
\includegraphics{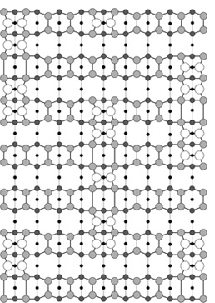} (b)
\includegraphics{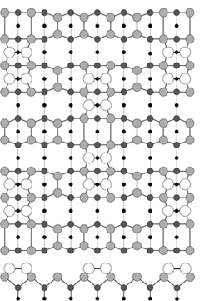} (c)
\caption{\label{fig:schematic} A schematic drawing of the $(8\times n)$ structure:  $(8\times 8)$ with the short blocks (a), a unit cell is outlined; the same structure with the long block (b);  $(8\times 6)$ structure  (c).}
\end{figure}

After cooling, the samples were moved into the STM chamber in which the pressure did not exceed  $10^{-10}$ Torr. The STM tip was {\it ex situ} made of the W wire and cleaned by ion bombardment\cite{W-tip}  
in a special UHV chamber connected to the STM chamber. 

The images were obtained in the constant tunneling current mode at room temperature. The STM tip was zero-biased while a sample was positively or negatively biased  for empty or filled states mapping.


Fig.~\ref{fig:view} demonstrates images of the surface prepared according to the above procedure. Steps with a height of $\sim 1.4$~\r{A} are seen in the pictures.  The surface is composed of rows running along $[110]$ and $[1\overline{1}0]$ directions. The areas marked as A, B and C lie on different terraces. The   area B is by $\sim 2.8$~\r{A} lower than the A one. In the regions A and B the rows are directed along the  $[1\overline{1}0]$ axis. A smaller domain with the rows directed along $[110]$ which lies between these regions is situated one step lower than the  region A. In the  region  C which is one step lower than the area B, the rows run along the $[110]$ axis.  So the rows which form the terraces separated by an even number of steps are parallel whereas the rows forming the terraces separated by an odd number of steps are  directed perpendicularly.

Each row consists of  rectangular  blocks which may be considered as structural units. 
Reflexes of the Fourier transform of the picture shown in Fig.~\ref{fig:view}(a)
correspond to the distances $\sim 31$~\r{A} and $\sim 15$~\r{A} in both $[110]$ and $[1\overline{1}0]$ directions (Fig.~\ref{fig:view}(c)). So the structure revealed in the long shot seems to have a periodicity of $\sim 31$~\r{A} that corresponds to 8 translations $a$ on the surface lattice of Si(001)  ($a=3.83$~\r{A} is a unit translation length). It looks like the Si(001)$-c(8\times 8)$ surface.\cite{Murray} STM images scanned at higher magnifications give an evidence that the surface appears to be disordered, though.


Fig.~\ref{fig:8x8} shows the magnified images. The rows of the   blocks are seen to be situated at varying distances from one another (hereinafter, the distances are measured between corresponding maxima of features). A unit $(8\times 8)$ cell is marked with a square box in Fig.~\ref{fig:8x8}(a). The distances between the adjacent rows of the  rectangles  are $4a$
in such structures. Fig.~\ref{fig:8x8}(b) demonstrates a surface area which contains a different structure of the rows. The adjacent rows designated as A are $3a$ apart. The rows marked as B go at $4a$ apart. Hence it may be concluded that order and some periodicity take place only along the rows. The structure is disordered across the rows. 

A structure with the rows going at $4a$ apart is presented in Fig.~\ref{fig:8x8}(c). The lost blocks looking like point defects are observed. In addition, a row wedging between two rows and as if separating them by an additional distance $a$ (the total distance becomes $5a$) is seen in the middle upper side of the picture.

Fig.~\ref{fig:empty-filled} demonstrates the empty  and filled  state images of the same surface are. Each block consists of two maxima clearly seen in both images. Fig.~\ref{fig:empty-filled}(c) shows the profiles of the images taken along the white lines. Extreme positions of both curves are  well fitted. Relative heights of the features outside and inside the blocks can be estimated from the profiles.

Before considering a model of the observed surface structure, let us dwell  on a close-up of the surface in more detail and especially on the  blocks as its unit elements.


Looking at the above images one can see  two types of blocks forming the surface structure---\,$\sim 15$~\r{A}  ($4a$) and $\sim 23$~\r{A} ($6a$) long. The distance between equivalent positions of the adjacent short blocks in the rows is $8a$. If the long block appears in a row a vacancy is formed in the adjacent row. Fig.~\ref{fig:close-up} illustrates this peculiarity---a short block is moved along the row out of its normal position by $2a$ because of the presence of the long one. Two vacancies and three long blocks are seen in Fig.~\ref{fig:close-up}(a). 
With this, a Fourier transform pattern looks like that for the undisturbed $c(8\times 8)$ structure (Fig.~\ref{fig:close-up}(b)), so the mentioned irregularities likely cannot be detected by means of integral techniques, e.g. by  LEED. In average the structure resembles the  $c(8\times 8)$ one.

It was found then that the  rectangles  are elevated over the surface through a height of at least 1.4~\r{A} that equals to the height of a monoatomic step on the Si(001) surface. A 3-D view of the area shown in the picture (a) illustrates it in Fig.~\ref{fig:close-up}(b). Measurements made on a number of STM images enabled the determination of all possible differences in height---within the blocks (both short or long), in gaps between adjacent blocks in a row and between neighboring rows, etc. So a surface relief was determined with an accuracy provided by the STM.

In addition, the long blocks were found to have one more peculiarity. They have extra maxima in their central regions. The maxima are not so pronounced as the main ones but nevertheless they are quite recognizable in the pictures (a) and (b) (Fig.~\ref{fig:close-up}).

Fig.~\ref{fig:block}(a) 
shows a close-up of the blocks giving an evidence that each  rectangle  consists of two rows separated by the distance close to $a$. Corresponding features are seen in Fig.~\ref{fig:block}(b) as well and marked with a circle and solid arrows. Precise measurement of this distance is difficult because an STM image reflects the electron density distribution of atoms situated both in the outermost layer and in the layer stretching under it.\cite{theor_sts} A total signal comprises a superposition of signals from several close atoms that sometimes results in line widening or displacement in the STM images. 

And at last, an STM signal is registered in the gaps between the neighboring rows  (Fig.~\ref{fig:block}(b), indicated with the dashed arrows). This signal is likely due to the atoms of the underlying layer.  Analogous features are also seen in Fig.~\ref{fig:8x8}.


The above data allowed us to draw a model of the observed Si(001) surface reconstruction.  This model takes into account the data of previous investigations carried out by different authors who suppose the  $c(8\times 8)$ structure to arise because of the presence of Cu atoms on the surface
\cite{Murray,Liu1,Liu2} as well as the results on low temperature deoxidation of the Si(001) surface according to which such process gives rise to $c(8\times 8)$ structure formation as well.
\cite{Ney}

The model is based on the following assumptions: (i) the outermost surface layer is formed by ad-dimers; (ii) the underlying layer has a structure of $(2\times 1)$; (iii) every  rectangular  block consists of ad-dimers a number of which controls the block length.

Fig.~\ref{fig:schematic}(a) shows a schematic drawing of the $(8\times 8)$-type structure (a unit cell is outlined). This structure is a basic one for the model brought forward. The elementary structural unit is a short  rectangle . These blocks form raised rows running vertically (shown by empty circles). Smaller shaded circles show horizontal dimer rows of the lower terrace. The rest black circles show bulk atoms. Each  rectangle  consists of two couples of dimers separated with a dimer vacancy. The structures on the Si(001) surface composed of close ad-dimers are known to be stable.\cite{Liu2} 
In our model, a position of the  rectangles  is governed by the location of the dimer rows of the $(2\times 1)$ structure of the underlying layer. The rows of blocks are always normal to the dimer rows in the underlying layer. Every  rectangular  block is bounded by the dimer rows of the underlying layer from both short sides. 

Fig.~\ref{fig:schematic}(b) demonstrates the same model for the case of the long  rectangle. This block is formed due to the presence of an additional dimer in the middle of the  rectangle. The structure consisting of one dimer is metastable,\cite{Liu2} so this type of blocks cannot be dominating in the structure. Each long block is bounded on both short sides by the dimer rows of the underlying terrace, too. The presence of the long  rectangle  results in the formation of a vacancy defect in the adjacent row, this is shown in Fig.~\ref{fig:schematic}(b)---the long  block  is drawn in the middle row, the dimer vacancy is present in the last left row.

\begin{figure}
\includegraphics{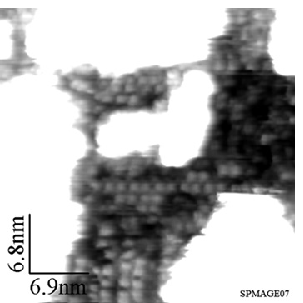} (a)
\includegraphics{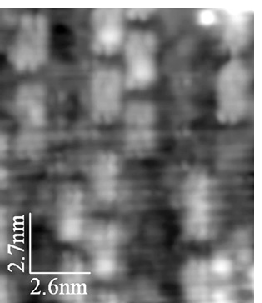} (b)
\caption{\label{fig:formation}Images of Si(001) after different deoxidation cycles in the UHV  MBE chamber:  annealing at $\sim 900\,^\circ$C  for $\sim 1.5$~min. (deoxidized partly) (a); effect of a faint flow of Si atoms at $\sim\,770\,^\circ$C (b);  +1.5~V, 150~pA and +1.9~V, 200~pA, respectively.}
\end{figure}

According to our STM data the surface is disordered in the direction perpendicular to the rows of the blocks. The distances between the neighboring rows may be less than those in the $(8\times 8)$ structure. Hence the structure presented in this paper may be classified as $(8\times n)$ one. Fig.~\ref{fig:schematic}(c) demonstrates an example of such a structure---a $(8\times 6)$ one. Formation of the $(8\times n)$ structure is explained by the diffusion of ad-dimers on the surface, and the diffusion along the dimer rows of the underlying layer being easier than across them.


An origin of this structure is not clear thus far. In the previous works studied the $c(8\times 8)$ structure,  authors attributed it to surface contamination by foreign atoms, e.g. by Cu.\cite{Hu,Murray,Kubo,Liu1,Ney} This seems rather probable but there is a circumstance that to some extent contradicts to this viewpoint.  A surface shown in Fig.~\ref{fig:formation}(a) was subjected to a shorter annealing than that applied in this work. It is deoxidized only in part. A surface shown in Fig.~\ref{fig:formation}(b) was deoxidized following a procedure different from that described in  this paper. A flux of Si atoms on the surface was applied and the sample temperature did not exceed $770\,^\circ$C.\cite{Report_01-303} Nevertheless, the surface structure similar to that described in the current paper was formed on the both samples. The effect of the surface contamination by foreign atoms cannot be completely excluded, though.


In summary, it may be concluded that the Si(001) surface prepared under the conditions of the UHV MBE chamber in a standard wafer preparation cycle has $(8\times n)$ reconstruction which is partly ordered only in one direction. Two types of unit blocks form the rows running along $[110]$ and $[1\overline{1}0]$ axes. When the long block disturbs the order in a row a dimer-vacancy defect appears in the adjacent row in the vicinity of the long block to restore the chess-board order of blocks in the neighboring rows.

The research was funded by the Science and Innovations Agency RF under the contract No.\,02.513.11.3130. 
The authors thank the developers of the free WSxM software\cite{WSxM} 
 which helped prepare the STM images.


\end{document}